\documentclass[prb,preprint,onlinecite,superscriptaddress,nobibnotes]{revtex4}

\usepackage{times}
\usepackage[pdftex]{graphicx}
\usepackage[pdftex]{epsfig}
\usepackage{amsmath}
\usepackage{pslatex}
\usepackage{amssymb}
\usepackage{amsfonts}
\usepackage{color}
\usepackage{wrapfig}
\usepackage{eucal}
\usepackage{hhline}
\usepackage{threeparttable}
\usepackage{supertabular}
\usepackage{multirow}
\usepackage{tabularx}
\usepackage{hyperref}

\usepackage{graphicx}
\usepackage{dcolumn}
\usepackage{bm}
\usepackage{times,mathptmx}
\usepackage{color}

\newcommand{\ave}[1]{\langle #1 \rangle}%
\newcommand{\dint} {\int\!\!\!\!\!\int}

\newcolumntype{Y}{>{\centering\arraybackslash}X}

\begin{document}

\pagenumbering{arabic}

\title{\bf Mechanical oscillation and cooling actuated by the optical gradient force}

\author{Qiang Lin, Jessie Rosenberg, Xiaoshun Jiang, Kerry J. Vahala, and Oskar Painter}
\email{opainter@caltech.edu}
\homepage{http://copilot.caltech.edu}
\affiliation{Thomas J. Watson, Sr., Laboratory of Applied Physics, California Institute of Technology, Pasadena, CA 91125}%

\date{\today}

\begin{abstract}
In this work we combine the large per-photon optical gradient force with the sensitive feedback of a high quality factor whispering-gallery microcavity.  The cavity geometry, consisting of a pair of silica disks separated by a nanoscale gap, shows extremely strong dynamical backaction, powerful enough to excite giant coherent oscillations even under heavily damped conditions (mechanical $Q \approx 4$).  In vacuum, the threshold for regenerative mechanical oscillation is lowered to an optical input power of only $270$ nanoWatts, or roughly $1000$ stored cavity photons, and efficient cooling of the mechanical motion is obtained with a temperature compression factor of $13$ dB for $4$ microWatt of dropped optical input power.   
\end{abstract}

\maketitle

Many precision position measurement devices involve the coupling of mechanical degress of freedom to an electromagnetic interferometer or cavity\cite{Braginsky77,Caves80}.  Today, cavity-mechanical systems span a wide range of geometries and scales, from multi-kilometer long gravitational-wave detectors\cite{Abramovici92} to coupled nanomechanical-microwave circuits\cite{Regal08}.  For the sensitive detection and actuation of mechanical motion, each of these systems depend upon ``dynamical backaction''\cite{Braginsky92,Kippenberg082} resulting from the position-dependent feedback of electromagnetic wave momentum.  Recent work in the optical domain has used the scattering radiation pressure force to both excite and dampen oscillations of a micro-mechanical resonator\cite{Kippenberg05,Gigan06,Arcizet06,Kleckner06,Schliesser06,ThompsonJD08,Schliesser08}, with the intriguing possibility of self-cooling the mechanical system down to its quantum ground-state.  

As has been recently proposed\cite{Povinelli052,Rakich07} and demonstrated\cite{Li08,Eichenfield09}, the optical gradient force within guided-wave nanostructures can be orders-of-magnitude larger than the scattering force.  In this work we combine the large per-photon optical gradient force with the sensitive feedback of a high quality factor whispering-gallery microcavity.  The cavity geometry, consisting of a pair of silica disks separated by a nanoscale gap, shows extremely strong dynamical backaction, powerful enough to excite giant coherent oscillations even under heavily damped conditions (mechanical $Q \approx 4$).  In vacuum, the threshold for regenerative mechanical oscillation is lowered to an optical input power of only $270$ nanoWatts, or roughly $1000$ stored cavity photons, and efficient cooling of the mechanical motion is obtained with a temperature compression factor of $13$ dB for $4$ microWatt of dropped optical input power.  These properties of the double-disk resonator make it interesting for a broad range of applications from sensitive force and mass detection in viscous environments such as those found in biology\cite{Horber03,Arlett06}, to quantum cavity-optomechanics in which a versatile, chip-scale platform for studying the quantum properties of the system may be envisioned.

\begin{figure}[ht]
\includegraphics[width=0.75\columnwidth]{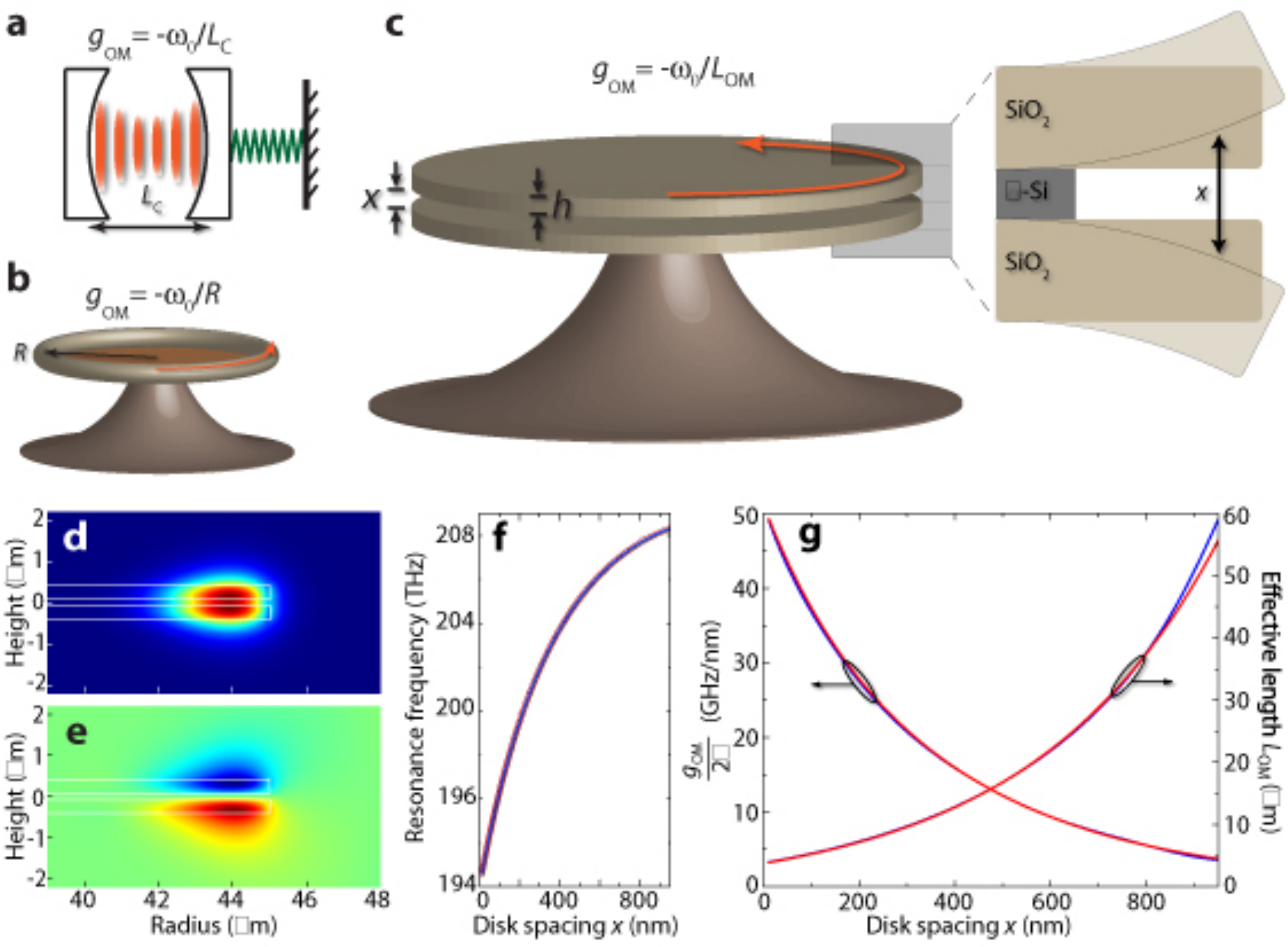}
\caption{\label{Fig1} Schematic of the corresponding (a) Fabry-Parot and (b) microtoroid optomechanical cavities. (c) Schematic of the double-disk NOMS structure, showing the mechanical flapping motion of the disks.  FEM-simulated optical mode profiles of the radial component of the electric field for the (d) bonded mode at $\lambda=1520$ nm and (e) antibonded mode at $\lambda=1297.3$ nm. (f) FEM-simulated tuning curve of the bonded mode.  (g) Optomechanical coupling coefficient and effective length (blue curves) for the bonded mode.  $g_{\text{OM}}$ and $L_{\text{OM}}$ are both well-approximated by exponential functions (red curves).}
\end{figure}

The per photon force exerted on a mechanical object coupled to the optical field within a resonant cavity is given by $\hbar g_{\text{OM}}$, where $g_{\text{OM}} \equiv \text{d}\omega_{c}/\text{d}x$ is a coefficient characterizing the dispersive nature of the cavity with respect to mechanical displacement, $x$.  In a Fabry-Perot (Fig.~\ref{Fig1}a) or microtoroid resonator (Fig.~\ref{Fig1}b), the optical force manifests itself as a so-called scattering radiation pressure due to direct momentum transfer from the reflection of photons at the cavity boundary\cite{Meystre85,Kippenberg07}. As the momentum change of a photon per round trip is fixed inside such cavities, while the round-trip time increases linearly with the cavity length, the radiation pressure per photon scales inversely with the cavity size.  In contrast, for the gradient optical force the cavity length and the optomechanical coupling can be decoupled, allowing for photon momentum to be transfered over a length scale approaching the wavelength of light\cite{Povinelli052,Rakich07}.  This method was recently employed in a silicon photonic circuit to manipulate a suspended waveguide\cite{Li08}.  However, without the feedback provided by an optical cavity or interferometer, the optical force only provides a static mechanical displacement.  

In the case of a cavity optomechanical system, dynamical backaction can be quantified by considering the magnitude of the damping/amplification that an input laser has on the mechanical motion.  For a fixed absorbed optical input power in the bad-cavity limit ($\kappa \gg \Omega_{m}$), the maximum rate is given by 

\begin{equation}
\label{eq:damp_coeff}
\Gamma_{m,\text{opt}} \approx \left(\frac{3\sqrt{3}g_{\text{OM}}^2}{\kappa_{i}^3\omega_{c}m_{x}}\right)\left(\frac{1}{(1+K)^2} \right)P_{d},
\end{equation}

\noindent where $\omega_{c}$ is the optical cavity resonance frequency, $m_{x}$ is the motional mass of the optomechanical system, $P_{d}$ is the optical power dropped (absorbed) within the cavity, and $K\equiv\kappa_{e}/\kappa_{i}$ is a cavity loading parameter ($\kappa_{i}$, the intrinsic energy loss rate of the optical cavity; $\kappa_{e}$, the energy coupling rate between external laser and internal cavity fields).  The effectiveness of the coupling between the optical and mechanical degrees of freedom can thus be described by a back-action parameter, $B = g_{\text{OM}}^2/\left(\kappa_{i}^3\omega_{c}m_{x}\right)$, which depends upon the motional mass, the per-photon force, and the optical cavity $Q$-factor.         

\begin{figure}[btp]
\includegraphics[width=0.75\columnwidth]{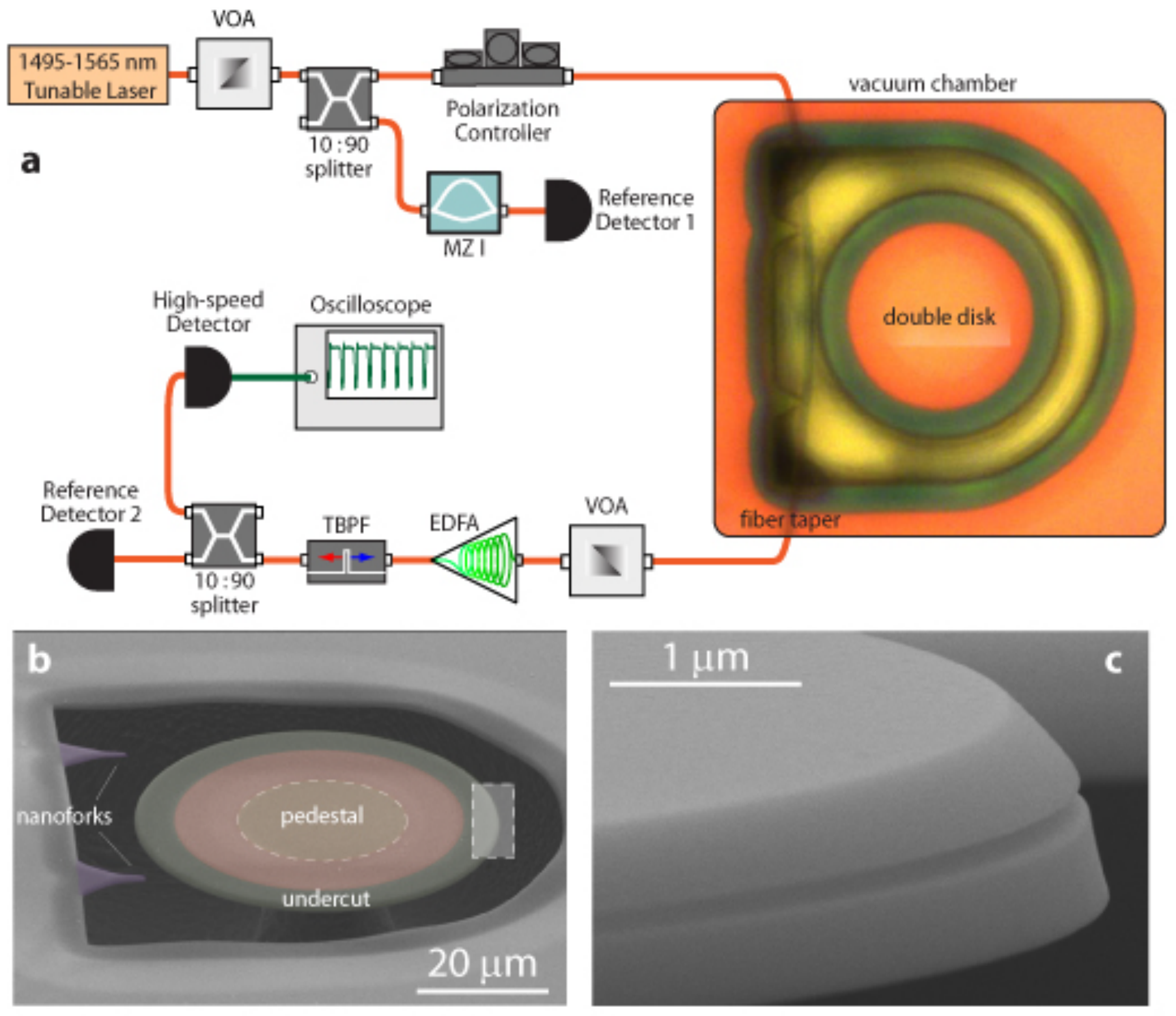}
\caption{\label{Fig2}(a) Schematic of the experimental setup for optical testing of the double-disk cavity. The cavity input and transmission are both transported through a single-mode silica fiber taper, which is supported by two nanoforks for stable operation. A tunable laser source is used to optically probe and actuate the double-disk structure, with input power controlled by a variable optical attenuator (VOA) and wavelength calibrated by a Mach-Zehnder interferometer (MZI). For experiments performed in a nitrogen environment, the cavity transmission is sent directly to the photodetectors, while it is first amplified by an erbium-doped fiber amplifier (EDFA) for the experiments performed in vacuum. (b-c) Scanning electron microscope images of the 54-${\rm \mu}$m double-disk NOMS. False color is used to indicate different relevant regions of the device.}
\end{figure}

Here we describe the design, fabrication, and characterization of a nano-optomechanical system (NOMS) consisting of a pair of optically thin disks separated by a nanoscale gap.  The double-disk structure (Fig.~\ref{Fig1}c) supports high-$Q$ whispering-gallery resonances, and provides back-action several orders of magnitude larger than in previously demonstrated gradient force optomechanical systems\cite{Li08,Eichenfield09} (very recent work\cite{Anetsberger09} involving the versatile coupling of external nanomechanical elements to the near-field of a high-Finesse microtoroid has realized very strong dynamical back-action, although still roughly two-orders of magnitude smaller than in our integrated device).  The double-disk cavities are formed a silicon wafer on which a multi-layer stack has been deposited consisting of silicon dioxide disk layers separated by a sacrificial amorphous silicon ($\alpha$-Si) layer.  Formation of the circular disk shape along with supporting fork structures are defined by electron-beam lithography and an optimized plasma dry-etch.  A dry release of the silica disk layers is performed using a highly-selective plasma etch of the $\alpha$-Si intermediate layer and the underlying Si substrate (see Methods).  The final double-disk structure, shown in Fig. \ref{Fig2}, consists of $340$-nm-thick silica disks separated by a 140 nm air gap extending approximately $6$ $\mu$m in from the disk perimeter (the undercut region).  Two different sized cavities are studied here, one large ($D=90$ $\mu$m; Sample I) and one small ($D=54$ $\mu$m; Sample II) in diameter.  The small diameter cavity structure represents a minimal cavity size, beyond which radiation loss becomes appreciable ($Q_{r} \sim 10^8$).      

Finite element method (FEM) simulations of the whispering-gallery optical modes of the double-disk structure shows substantial splitting of the cavity modes into even and odd parity bonded and anti-bonded modes (Fig.~\ref{Fig1}e-f).  Due to its substantial field intensity within the air gap, the bonded mode tunes rapidly with changing gap size as shown in the inset to Fig. \ref{Fig1}g.  The optomechanical coupling coefficient, $g_{\text{OM}}$, defined as the derivative of the mode tuning curve, can be related to an effective optomechanical coupling length, $L_{\text{OM}}$ through the relation $g_{\text{OM}} \equiv \omega_{c}/L_{\text{OM}}$.  As the two disks are coupled through the evanescent field between them, $L_{\text{OM}}$ decreases exponentially with disk spacing (Fig.~\ref{Fig1}g), reaching a minimum value of $3.8$ $\mu$m at a resonance optical wavelength of $\lambda_{c} \approx 1.5$ $\mu$m. For the air gap of $138$ nm used in this work, the optomechanical coupling is estimated to be $g_{\text{OM}}/2\pi = 33$ GHz/nm ($L_{\text{OM}}=5.8$ $\mu$m), equivalent to $22$ fN/photon. 

The double-disk structure also supports a number of different micro-mechanical resonances, ranging from radial breathing modes to whispering-gallery-like vibrations of the disk perimeter.  The most strongly coupled mechanical resonance is that of the symmetric (i.e., azimuthal mode number, $m=0$) flapping motion of the disks.  Due to the symmetry of the gradient force on the two disks, only the 6 $\mu$m undercut air gap region is involved in the flapping motion (Fig.~\ref{Fig1}c and Fig.~\ref{Fig2}), with negligible displacement of the interior of the disk. Consequently, the flapping mode exhibits an effective motional mass, $m_{x}$, of only $145$ and $264$ picrograms for the $54$ and $90$ $\mu$m diameter cavities, respectively (see App. \ref{appA}).  Note that both these values are more than two orders of magnitude smaller than commonly used micromirrors and microtoroids \cite{Meystre85,Gigan06,Arcizet06,Kleckner06,Schliesser06,ThompsonJD08,Schliesser08}, and in combination with the large per-photon force, provide a significant enhancement to the dynamic back-action parameter which scales as $g_{\text{OM}}^2/m_{x}$.

\begin{figure}[btp]
\includegraphics[width=0.75\columnwidth]{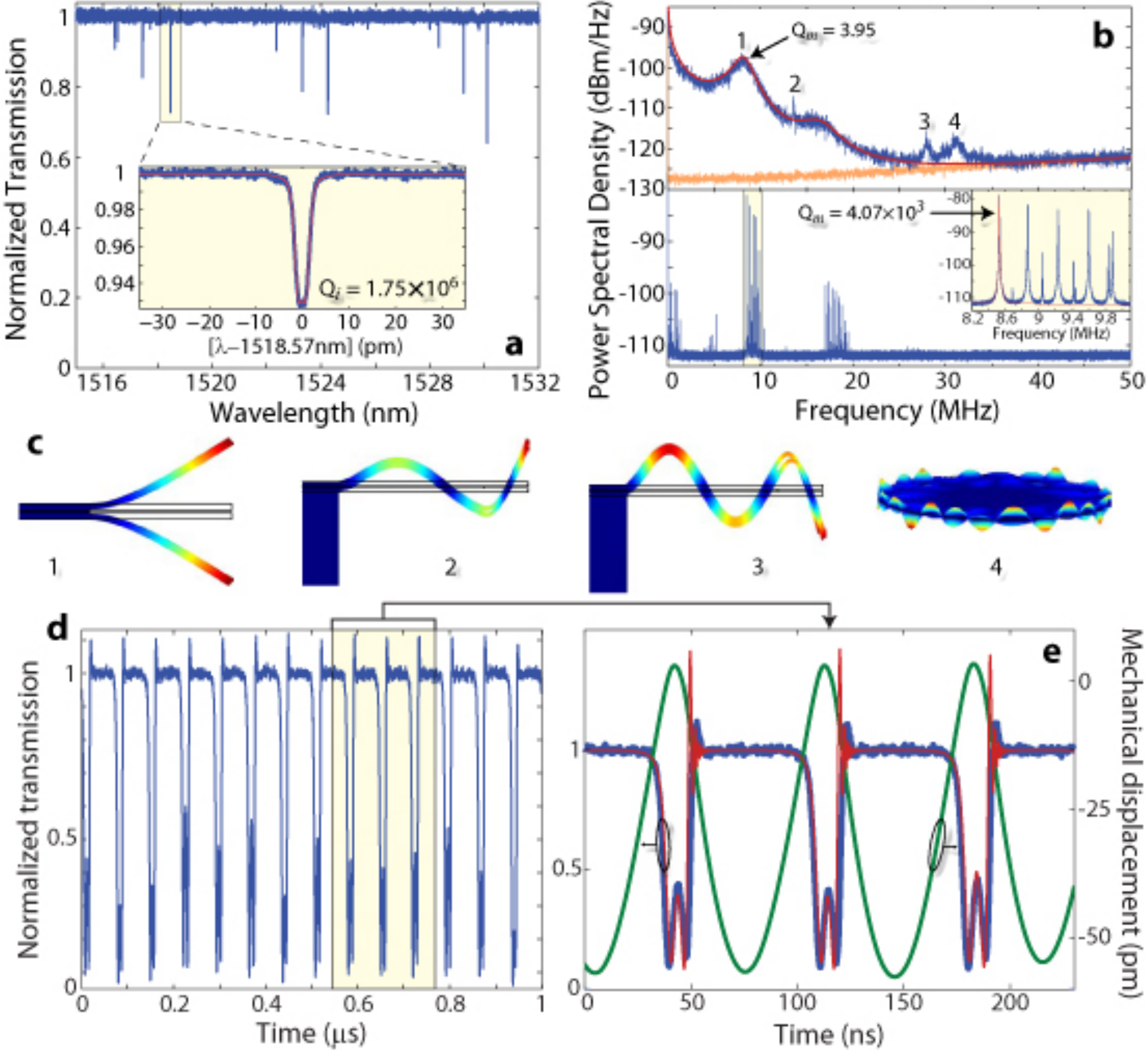}
\caption{\label{Fig3}(a) Optical transmission spectrum of a large diameter ($D=90$ $\mu$m; Sample I) double-disk cavity. The inset shows the fundamental TE-like bonded mode at $\lambda=1518.57$ nm. (b) Upper panel: optical transmission power spectral density (PSD) of a Sample I double-disk in the 1 atm. nitrogen environment for $P_{i}=5.8$ $\mu$W.  Experimental data in blue, theoretical modeling in red, and detector noise background in yellow. Lower panel: transmission PSD of a small diameter ($D=54$ $\mu$m; Sample II) double-disk cavity in vacuum for $P_{i}=44$ nW. The inset shows a zoomed-in of the spectrum around the fundamental flapping mode frequency. (c) FEM simulated mechanical modes indicated in (b). (d) Recorded transmission waveform of Sample I for $P_{i}=0.76$ mW. (e) Comparison of experimental (blue curve) and simulated (red curve) waveforms, with the corresponding simulated mechanical displacement (green curve). Further details of the measurement conditions are given in App. \ref{appC}.}
\end{figure}

Optical and mechanical measurements were initially performed at room temperature in a one atmosphere nitrogen environment.  Fig. \ref{Fig3}a shows the wavelength scan of a large diameter double-disk cavity (Sample I).  Several radial-order whispering-gallery modes are evident in the spectrum, all of them of TE-like polarization and bonded mode character.  The fundamental TE-like bonded optical mode at $\lambda=1518.57$ nm is shown in the Fig. \ref{Fig3}a inset, from which an intrinsic optical $Q$-factor of $1.75 \times 10^6$ is inferred (see App. \ref{appB}). The radio-frequency (RF) power spectrum of the optical signal transmitted through the cavity (Fig.~\ref{Fig3}b, top panel) exhibits three clear frequency components at $8.30$, $13.6$, and $27.9$ MHz corresponding to thermally-actuated resonances of the double-disk structure. These values agree well with FEM simulations of the differential flapping mode mode ($7.95$ MHz), and the first ($14.2$ MHz) and second ($28.7$ MHz) order radial breathing modes (Fig.~\ref{Fig3}c). The strong dynamic back-action of the flapping mode (under thermal excitation) also produces a broadband spectral background in the RF spectrum with a shoulder at the second harmonic frequency (see App. \ref{appB}). The correct description of the power spectrum (Fig.~\ref{Fig3}b, red curve) shows that the flapping mode has a 3-dB linewidth of $2.1$ MHz (mechanical $Q$-factor, $Q_{M}=3.95$), limited by the squeeze-film process of the nitrogen gas between the disks\cite{Verbridge082}.

Despite the near-unity mechanical quality factor of the flapping mode, the powerful dynamic back-action in the double-disk structure provides sufficient compensation of mechanical loss to excite regenerative mechanical oscillation. As shown in Fig.~\ref{Fig3}d, with an input optical power of $760$ $\mu$W launched at the blue detuned side of the resonance, the induced parametric mechanical instability causes the cavity transmission to oscillate over the entire coupling depth with a fundamental frequency of $13.97$ MHz (this value is about 68\% larger than the intrinsic mechanical frequency due to the optical spring effect\cite{Sheard04}). A zoom-in of the recorded time waveform (Fig.~\ref{Fig3}e) agrees well with our numerical simulation which shows that the gradient force actuates an extremely large ($50$ pm) mechanical displacement amplitude, dragging the cavity resonance over more than 10 cavity-linewidths and leaving distinctive features of the Lorentzian cavity transfer function.  In particular, two sequential passes of the cavity resonance across the laser frequency can be seen, along with an overshoot and oscillation of the transmitted optical power resulting from the quick release of Doppler shifted photons from the cavity.

\begin{figure}[btp]
\includegraphics[width=0.75\columnwidth]{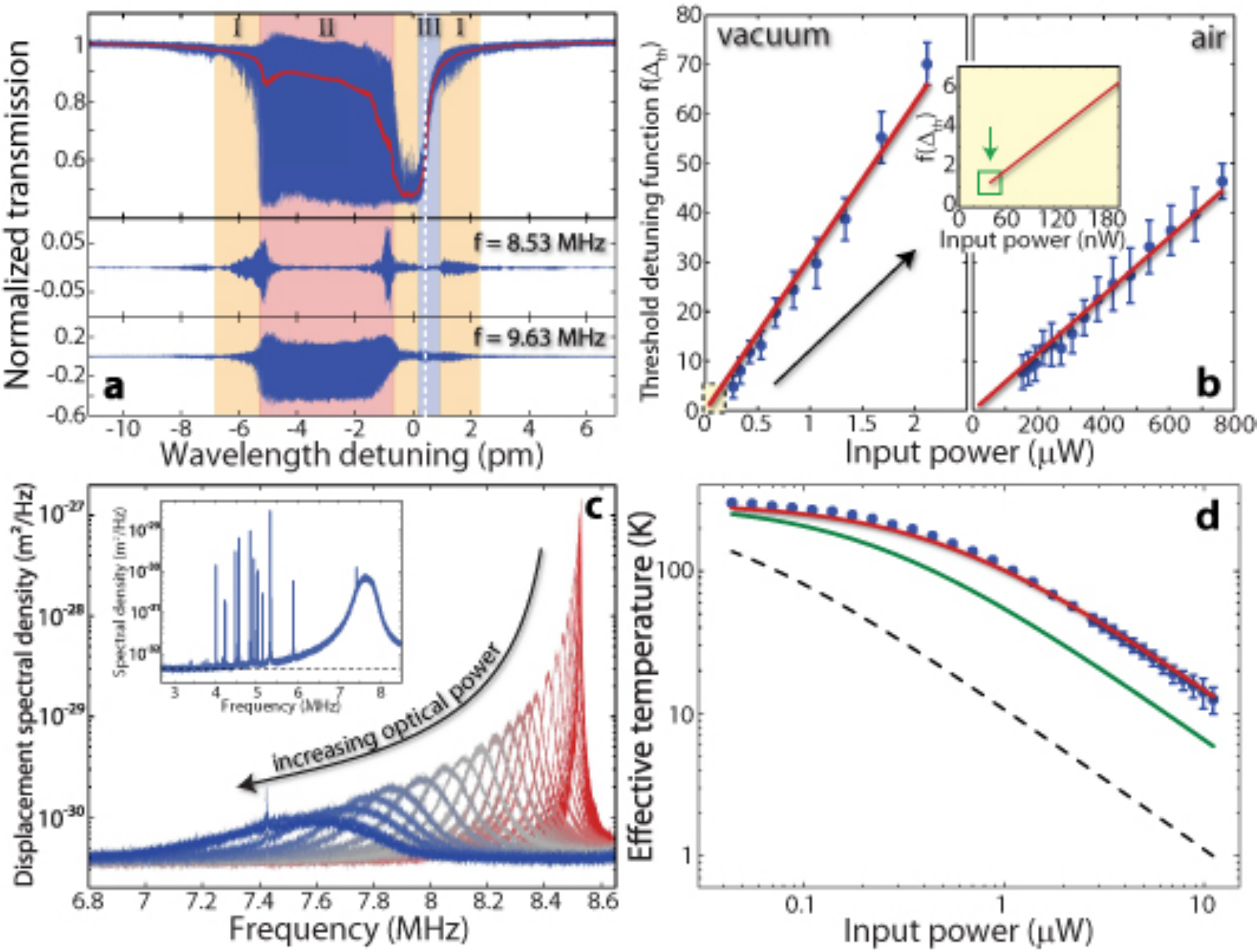}
\caption{\label{Fig4}(a) Top panel: Normalized cavity transmission for Sample II in vacuum and $P_{i}=11$ $\mu$W. Blue and red traces show the instantaneous and low-pass-filtered signals, respectively. Middle panel: the transduction amplitude of the frequency component at $8.53$ MHz and its higher-order harmonics. Bottom panel: the transduction amplitude of the frequency component at $9.63$ MHz and its higher-order harmonics. (b) $f(\Delta_{\rm th})$ as a function of optical input power.  Right Panel: Sample I in a 1 atm. nitrogen environment. Left panel: Sample II in vacuum (inset shows the minimum achievable threshold (green arrow)). (c) Spectral intensity of the thermally-driven fundamental flapping mode at various input powers, recorded for Sample II in vacuum, with a laser detuning of $\Delta=-1.45(\kappa/2)$ (inset shows the displacement sensitivity at the highest input power with the second optical attenuator removed), and (d) the corresponding effective temperature. In (d) the red curve is a fit to the data, the solid green (dashed black) curve is a theoretical curve obtained using the estimated $B$-parameter from the left panel of (b) and the experimental (optimal) detuning of $\Delta=-1.45(\kappa/2)$ $\left(\Delta=-(\kappa/2)/\sqrt{5}\right)$.}
\end{figure}

The threshold for regenerative oscillation depends sensitively upon the optical input power and the average laser-cavity resonance detuning, a map of which can be used to quantify the strength of the dynamic back-action.  An estimate of the threshold detuning ($\Delta_{\rm th}$), for a given input power, can be determined from the abrupt kink in the cavity transmission that marks the onset of regenerative oscillation (Fig.~\ref{Fig4}a and App. \ref{appC}).  The detuning dependence of the optomechanical amplification coefficient can be lumped into a single detuning function, 

\begin{equation}
f(\Delta) \equiv \left(\frac{ \Delta^2+ (\kappa/2)^2 }{ \kappa \kappa_e \kappa_i^3 \Delta}\right) \left( (\Delta + \Omega_m)^2 + \left( \frac{\kappa}{2} \right)^2 \right) \left( (\Delta - \Omega_m)^2 + \left( \frac{\kappa}{2} \right)^2 \right). \label{f_Delta}
\end{equation}

\noindent where $\kappa = \kappa_{i} + \kappa_{e}$ is the total photon decay rate of the loaded cavity.  The right panel of Fig.~\ref{Fig4}b shows a map of $f(\Delta_{\rm th})$ versus optical input power for the $90$ $\mu$m diameter double-disk cavity in the heavily damped nitrogen environment.  The data in Fig.~\ref{Fig4}b, as expected, shows a linear dependence of $f(\Delta_{\rm th})$ on input power, and is well described in the unresolved sideband regime\cite{Kippenberg07} by 

\begin{equation}
f(\Delta_{\rm th}) = \frac{2 g_{\text{OM}}^2 P_{i}}{\omega_{c} m_{x} \Gamma_{m} \kappa_i^3} = \left(\frac{2B}{\Gamma_{m}}\right)P_{i}, \label{Threshold}
\end{equation}

\noindent where $\Gamma_{m}=2.1$ MHz is the bare mechanical damping rate of the flapping mode. Fitting of eq. ~(\ref{Threshold}) to the data in Fig.~\ref{Fig4}b yields a dynamic back-action parameter of $B=0.061$ MHz/$\mu$W, corresponding to an optomechanical coupling factor of $g_{\text{OM}}/2\pi = 33.8 \pm 0.4$ GHz/nm, in good agreement with the simulated result of $33$ GHz/nm.

In order to eliminate the squeeze-film damping of the nitrogen environment, measurements were also performed in vacuum ($P < 5 \times 10^{-4}$~Torr).  The significantly reduced mechanical linewidth in vacuum shows that the flapping mode consists of a small cluster of modes (Fig.~\ref{Fig3}b, bottom panel).  As discussed more fully in App. \ref{appC}, these modes are a mixture of the lower-lying azimuthal modes, coupled together due to deviations in circularity of the undercut region and support pedestal.  Measurements of the optical spring effect indicates that the optical field renormalizes the cluster of modes, with the lowest-frequency mode at $8.53$ MHz transforming into the fundamental flapping mode with uniformly distributed displacement along the disk perimeter (the rest of the modes decouple from the light field).  With an in-vacuum $Q_{M}=4070$ (Fig.~\ref{Fig3}b, inset), the fundamental flapping mode has an extremely low threshold input power for regenerative oscillation.  Figure \ref{Fig4}a shows a transmission spectrum when the laser is scanned across the cavity resonance.  Three different regimes can be clearly seen: (I) transduction of thermal motion, (II) onset of optically-driven oscillation, and (III) optically damped motion.  The onset of regenerative oscillation coincides with a frequency shift in the fundamental flapping mode to $9.63$ MHz as shown in the bottom two panels of Fig.~\ref{Fig4}a.  The left panel of Figure~\ref{Fig4}b shows a plot of the in-vacuum $f(\Delta_{\rm th})$ versus input power, with a measured minimum threshold power of $P_{i}=267$ nW.  Extrapolation of the experimental data using Eqs.~(\ref{f_Delta}) and (\ref{Threshold}) to the optimal detuning point shows a minimum threshold power of only $40$ nW.

The large mechanical amplification of the double-disk NOMS implies a correspondingly efficient cooling of mechanical motion on the red-detuned side of the cavity resonance. As shown in Fig.~\ref{Fig4}c for Sample II in vacuum, the spectral intensity of the fundamental flapping mode decreases dramatically with increased input power, accompanied by a significant broadening of the mechanical linewidth.  Even for the strongest damping levels, the inset to Fig.~\ref{Fig4}c shows good signal to noise for the transduced motion due to the high displacement sensitivity of the double-disk ($7\times10^{-17}$ m/Hz$^{1/2}$, as limited by the background level).  A measure of the optical cooling can be determined from the integrated area under the displacement spectrum\cite{Cohadon99} (see App. \ref{appD}).  Figure ~\ref{Fig4}d plots the inferred temperaure, $T_{\rm eff}$, which drops down to $12.5$~K for a maximum input power of $P_{i}=11$ $\mu$W ($P_{d}=4.4$ $\mu$W).  In principle, the effective temperature is related to the optical damping rate ($\Gamma_{m,\text{opt}}$) through the relation $T_0/T_{\rm eff} = 1 + \Gamma_{m}/\Gamma_{m,\text{opt}}$, where $T_{0}=300$~K is the bath temperature.  In Fig.~\ref{Fig4}d the red curve is a fit of the measured cooling curve using the relation $T_0/T_{\rm eff} = 1 + \alpha P_{i}$, whereas the green curve represents the expected cooling curve for the dynamic back-action parameter ($B=0.032$ MHz/$\mu$W) determined from the threshold plot in the right panel of Fig.~\ref{Fig4}b and the experimental laser-cavity detuning ($\Delta = -1.45(\kappa/2)$).  For comparison, we have also plotted (dashed black line) the theoretical cooling curve in the case of optimal laser-cavity detuning ($\Delta = -(\kappa/2)/\sqrt{5}$).  The difference between the two theoretical curves and the measured data, along with the limited range of optical input power studied, can largely be attributed to issues associated with the limited bandwidth and range of our current cavity locking scheme (a problem exacerbated by the very large transduction of even the Brownian motion of the disks).  As the dashed black curve indicates, technical improvements in the cavity locking position and stability should enable temperature compression factors of $20$ dB for less than $1$ $\mu$W of dropped power.            

The large dynamic back-action of the double-disk cavity, primarily a result of the large per-photon force and small motional mass of the structure, opens up several areas of application outside the realm of more conventional ultra-high-$Q$ cavity geometries.  This can be seen by considering not only the efficiency of the cooling/amplification process, but also the maximum rate of effective cooling/amplification, the scale of which is set by the optical cavity decay rate\cite{Marquardt07,Wilson-Rae07}.  In the double-disk cavities presented here, the dynamic back-action parameter is $B\approx 0.06$ MHz/$\mu$W for a cavity decay rate of $\kappa/2\pi \approx 100$ MHz.  The combination allows for higher mechanical frequencies of operation, where the bare damping is expected to scale with frequency, and makes possible enormous temperature compression ratios.  A quantum mechanical analysis of the optical self-cooling process\cite{Marquardt07,Wilson-Rae07}, indicates that the sideband resolved regime ($\kappa \lesssim \sqrt{32}\Omega_{m}$) is necessary to reduce the phonon occupancy below unity.  Having already achieved optical $Q$-factors in excess of $10^6$, and planar silica microdisks having already been demonstrated with $Q > 10^7$ \cite{Kippenberg03}, we expect that further optimization of the double-disk NOMS will be able to extend its operation well into the sideband resolved regime.  The combination of large dynamic back-action parameter and large maximum amplification rate also present intriguing possibilites for sensitive, high temporal resolution force detection\cite{Stowe97}, particularly in heavily damped environments such as fluids for biological applications\cite{Horber03,Arlett06}.  Other application areas enabled by the chip-scale format of these devices include tunable photonics\cite{Povinelli052,Rakich07,Li08}, optical wavelength conversion\cite{Notomi06}, and RF-over-optical communication.




\section*{Acknowledgements} The authors would like to thank Patrick Herring and Matt Eichenfield for some of the early development of the double-disk structure, Thomas Johnson and Raviv Perahia for help with device processing, Ryan Camacho for help in automating some of the measurements, and Thiago Alegre for useful discussions pertaining to the data anlysis.  This work was supported by a DARPA seedling grant and an NSF EMT grant.   


\appendix

\section{Double-disk fabrication}
\label{app0}
Fabrication of the double-disk whispering-gallery resonator consisted of initial deposition of the cavity layers.  The two silica disk layers and the sandwiched amorphous silicon ($\alpha$-Si) layer were deposited on a (100) silicon substrate by plasma-enhanced chemical vapor deposition, with a thickness of $340 \pm 4$~nm and $158 \pm 3$~nm for the silica and $\alpha$-Si layers, respectively. The wafer was then thermally annealed in a nitrogen environment at a temperature of $T=1050$ K for 6 hours to drive out water and hydrogen in the film, improving the optical quality of the material. The disk pattern was created using electron beam lithography followed by an optimized C$_4$F$_8$-SF$_6$ gas chemistry reactive ion etch.   Release of the double disk structure was accomplished using a SF$_6$ chemical plasma etch which selectively ($30,000:1$) attacks the intermediate $\alpha$-Si layer and the underlying Si substrate, resulting in a uniform undercut region between the disks which extends radially inwards 6~${\rm \mu}$m from the disk perimeter.  Simultaneously, the underlying silicon support pedestal is formed. The final gap size between the disks was measured to be $138 \pm 8$~nm (shrinkage having occurred during the anneal step). Two nanoforks were also fabricated near the double-disk resonator to mechanical stabilize and support the fiber taper during optical coupling; the geometry was optimized such that the forks introduce a total insertion loss of only $\sim$8\%.

\section{Optical gradient force and effective motional mass}
\label{appA}
\subsection{Optical gradient force in a double-disk NOMS}
\label{appA1}
\begin{figure}[tbp]
\scalebox{0.80}[0.80]{\includegraphics{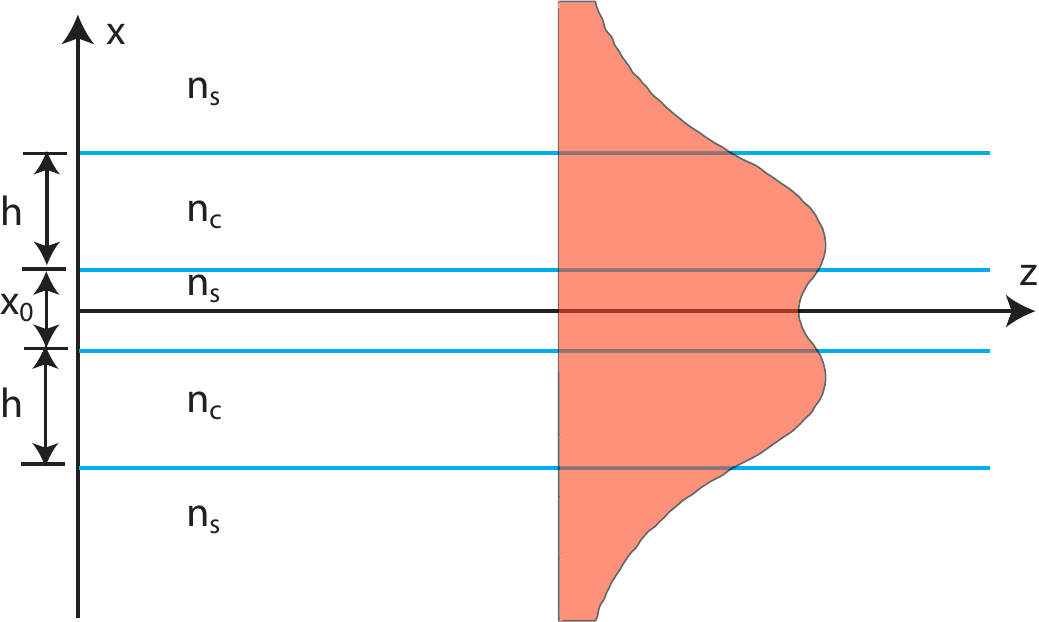}}
\caption{\label{SubFig_DoubleSlab} Schematic of a symmetric double-slab waveguide. $h$ and $x_0$ are the slab thickness and the slab spacing, respectively. $n_c$ and $n_s$ are the refractive indices for the slab and surrounding area, respectively. }
\end{figure}

As the mode confinement in a double-disk NOMS is primarily provided by the transverse boundaries formed by the two disks, the double-disk structure can be well approximated by a symmetric double-slab waveguide shown in Fig.~\ref{SubFig_DoubleSlab}. For the bonding mode polarized along the $\hat{e}_y$ direction, the tangential component of the electric field is given by:
\begin{eqnarray}
E_y = \left\{ {\begin{array}{*{20}c}
   {A e^{-\gamma x}, \qquad \qquad x > h+x_0/2}  \\
   {B \cos \kappa x + C \sin \kappa x, \qquad \qquad x_0/2 < x < h+x_0/2}  \\
   {D \cosh \gamma x, \qquad \qquad -x_0/2 < x < x_0/2}  \\
   {B \cos \kappa x - C \sin \kappa x, \qquad \qquad -x_0/2 > x > -h-x_0/2}  \\
   {A e^{\gamma x}, \qquad \qquad x < -h-x_0/2}  \\
\end{array}} \right. \label{Ey}
\end{eqnarray}
where $\kappa$ is the transverse component of the propagation constant inside the slabs and $\gamma$ is the field decay constant in the surrounding area. They are given by the following expressions:
\begin{eqnarray}
\kappa^2 = k_0^2 n_c^2 - \beta^2, \qquad \gamma^2 = \beta^2 - k_0^2 n_s^2, \label{k2_gamma2}
\end{eqnarray}
where $k_0=\omega_0/c$ is the propagation constant in vacuum and $\beta = k_0 n_{\rm eff}$ is the longitudinal component of the propagation constant of the bonding mode. $n_{\rm eff}$ is the effective refractive index for the guided mode. Accordingly, the tangential component of the magnetic field can be obtained through $H_z = \frac{-i}{\mu \omega_0} \frac{\partial E_y}{\partial x} $. The continuity of $E_y$ and $H_z$ across the boundaries requires $\kappa$ and $\gamma$ to satisfy the following equation:
\begin{equation}
\kappa \gamma \left[1+\tanh (\gamma x_0/2) \right] = \left[\kappa^2 - \gamma^2 \tanh (\gamma x_0/2) \right]\tan \kappa h , \label{BoundaryCondition}
\end{equation}
which reduces to $\tan \kappa h = \gamma/ \kappa$ when $x_0 \rightarrow 0$, as expected.

The circular geometry of the double disk forms the whispering-gallery mode, in which the resonance condition requires the longitudinal component of the propagation constant, $\beta$, to be fixed as $2 \pi R \beta = 2 m \pi$, where $R$ is the mode radius and $m$ is an integer. Thus, any variation on the disk spacing $x_0$ transfers to a variation on the resonance frequency $\omega_0$ through Eqs.~(\ref{k2_gamma2}) and (\ref{BoundaryCondition}), indicating that $\omega_0$ becomes a function of $x_0$. By using these two equations, we find that the optomechanical coupling coefficient, $g_{\text{OM}} = \frac{d\omega_0}{dx_0}$, is given by the general form 
\begin{eqnarray}
g_{\text{OM}}(x_0) = \frac{\frac{c \chi \gamma^2}{k_0} {\rm sech}^2\left(\frac{\gamma x_0}{2}\right)}{4 (n_c^2 - n_s^2) \tan \kappa h + n_s^2 x_0 \chi {\rm sech}^2 \left( \frac{\gamma x_0}{2} \right) + 2 \xi \left[ \left( n_c^2 \gamma h \csc^2 \kappa h + 2 n_s^2 \right) \tan \kappa h + \frac{n_s^2 \kappa}{\gamma} - \frac{n_c^2 \gamma} {\kappa} \right] }, \label{g_OM_2slab_general}
\end{eqnarray}
where $\chi \equiv \kappa + \gamma \tan \kappa h$ and $ \xi \equiv 1+ \tanh (\frac{\gamma x_0}{2})$.

When $x_0 \rightarrow 0$, Eq.~(\ref{g_OM_2slab_general}) leads to the maximum optomechanical coupling as
\begin{equation}
g_{\text{OM}}(0) = \frac{ \omega_0 \gamma^3}{2\beta^2 + 2 k_0^2 n_c^2 \gamma h}. \label{g_OM_Max}
\end{equation}
In analogy to Fabry-Parot cavities and microtoroids, the magnitude of the optomechanical coupling can be characterized by an effective length, $L_{\text{OM}}$, defined such that $g_{\text{OM}} \equiv \frac{\omega_0}{L_{\text{OM}}}$. Equation~(\ref{g_OM_Max}) infers a minimum effective length 
\begin{equation}
L_0= \frac{2}{\gamma} \left[ 1 + \frac{k_0^2}{\gamma^2}(n_s^2+n_c^2 \gamma h) \right] = \frac{\lambda_0}{\pi} \frac{n_{\rm eff}^2 + k_0 h n_c^2 \sqrt{n_{\rm eff}^2-n_s^2}}{\left(n_{\rm eff}^2-n_s^2 \right)^{3/2}}, \label{L_0}
\end{equation}
which is approximately on the order of the optical wavelength $\lambda_0$.

Physically, as the two slabs are coupled through the evanescent field between them with amplitude decaying exponentially with slab spacing at a rate $\gamma$ [see Eq.~(\ref{Ey})], the resulting optomechanical coupling can be well approximated by an exponential function 
\begin{equation}
g_{\text{OM}}(x_0) \approx g_{\text{OM}}(0) e^{-\gamma x_0}, \label{g_OM_approx}
\end{equation}
where $g_{\text{OM}}(0)$ is given by Eq.~(\ref{g_OM_Max}). As indicated by the red curve in Fig.~1(g) in the main text, Eq.~(\ref{g_OM_approx}) provides an excellent approximation for the optomechanical coupling coefficient in a double-disk NOMS. Therefore, the approximate effective length, $L_{\text{OM}} \approx \frac{\omega_0}{g_{\text{OM}}(0)} e^{\gamma x_0}$, agrees well with the results simulated by the finite element method [see Fig.~1(g) of the main text], and the effective length grows roughly exponentially with the disk spacing.

\subsection{Effective motional mass for the flapping mode}
\label{appA2}
\begin{figure}[tbp]
\scalebox{1.30}[1.30]{\includegraphics{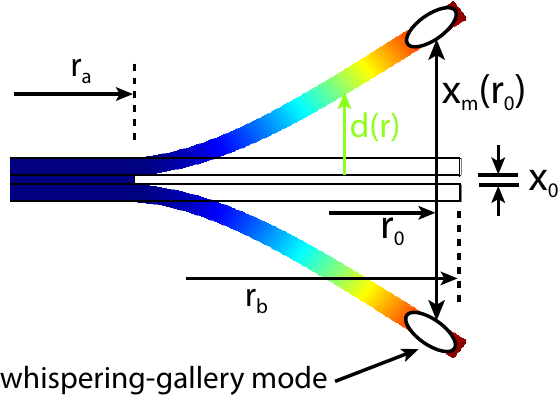}}
\caption{\label{SubFig_EffectiveMass} Illustration of the disk displacement. $x_0$ is the disk spacing in the absence of the optical field. $r_0$ is the radius of the whispering-gallery mode. $x_m(r_0)$ corresponds to the effective disk spacing at the mode center. $r_a$ and $r_b$ are the inner and outer radii of the disk region involved in the flapping motion. $d(r)$ is the mechanical displacement at radius $r$.}
\end{figure}

With a clamped inner edge and a free outer edge, the mechanical displacement of a double disk exhibiting a flapping mode is generally a function of radius (Fig.~\ref{SubFig_EffectiveMass}). What matters the optomechanical effect, however, is the disk spacing at the place where the whispering-gallery mode is located, as that determines the magnitude of the splitting between the bonding and antibonding cavity modes. As the mechanical displacement actuated by the gradient force is generally small compared with the original disk spacing $x_0$, we can assume it is uniform in the region of the whispering-gallery mode and define the effective disk spacing $x_m(r_0)$ at the mode center, where $r_0$ is the radius of the whispering-gallery mode. The effective mechanical displacement is then given by $x_{\rm eff} = x_m(r_0)-x_0$, corresponding to an effective mechanical potential energy of $E_p = m_{x} \Omega_m^2 x_{\rm eff}^2/2$, where $m_{x}$ is the corresponding effective motional mass and $\Omega_m$ is the resonance frequency of the flapping mode. Note that $x_{\rm eff}$ is twice as the real displacement at the mode center for a single disk, $x_{\rm eff}=2 d(r_0)$. $E_p$ reaches its maximum value when the double disk is at rest at its maximum displacement, at which point all of the mechanical energy is stored in the strain energy $U_s$. Therefore, $E_p = U_s$ and the effective motional mass is given by
\begin{equation}
m_{x} = \frac{ 2 U_s}{\Omega_m^2 [x_m(r_0)-x_0]^2} = \frac{ U_s}{2 \Omega_m^2 d^2(r_0)}, \label{m_eff}
\end{equation}
where both $U_s$ and $d(r_0)$ can be obtained from the mechanical simulations by the finite element method.

The relationship between the effective mass and the physical mass of the double-disk NOMS can be found by examining the mechanical potential energy. With a mechanical displacement $d(r)$ for each single disk [Fig.~\ref{SubFig_EffectiveMass}], we can find the total mechanical potential energy by integrating over the disk regions involved in the flapping motion:
\begin{equation}
E_p = \int_{r_a}^{r_b} {\Omega_m^2 d^2(r) \zeta 2 \pi r h dr}, \label{PotentialEnergy}
\end{equation}
where $\zeta$ is the material density, $h$ is the thickness for a single disk, $r_a$ and $r_b$ are the inner and outer radii of the disk region involved in the flapping motion (see Fig.~\ref{SubFig_EffectiveMass}). Note that $E_p$ is the total potential energy for the two disks, which is simply two times that of single one because of the symmetry between the two disks. As the physical mass of a single disk region involving in the flapping motion is given by $m_p = \pi \zeta h (r_b^2 - r_a^2)$, using Eq.~(\ref{PotentialEnergy}), we find that the effective mass is related to the physical mass through the following expression:
\begin{equation}
m_{x} = \frac{4 m_p}{\left( r_b^2 - r_a^2 \right) \left[ x_m(r_0)- x_0 \right]^2 } \int_{r_a}^{r_b} {r d^2(r) dr} =  \frac{m_p}{\left( r_b^2 - r_a^2 \right) d^2(r_0) } \int_{r_a}^{r_b} {r d^2(r) dr}. \label{Meff_Mp}
\end{equation}
As the whispering-gallery mode is generally located close to the disk edge (\emph{i.e.}, the mode radius $r_0=44~{\rm \mu m}$ in a double disk with $r_b = 45~{\rm \mu m}$), $d^2(r)/d^2(r_0) \ll 1$ for most of the region between $r_a$ and $r_b$, and Eq.~(\ref{Meff_Mp}) shows that $m_{x} \ll m_p/2$. Therefore, the effective mass is significantly less than half the physical mass of a single disk region. In practice, the effective mass is much smaller than this value because of the real displacement function $d(r)$. For the 90-$\mu$m device used in our experiment, the effective mass is 0.264 nanogram, only about one fifth of the physical mass of a single disk region $m_p = 1.18$~nanogram. The effective mass decreases to 0.145 nanogram for the 54-$\mu$m device, due to the decrease in the disk radius.

\section{Linear transmission of an optomechanical cavity}
\label{appB}
\begin{figure}[tbp]
\scalebox{0.70}[0.70]{\includegraphics{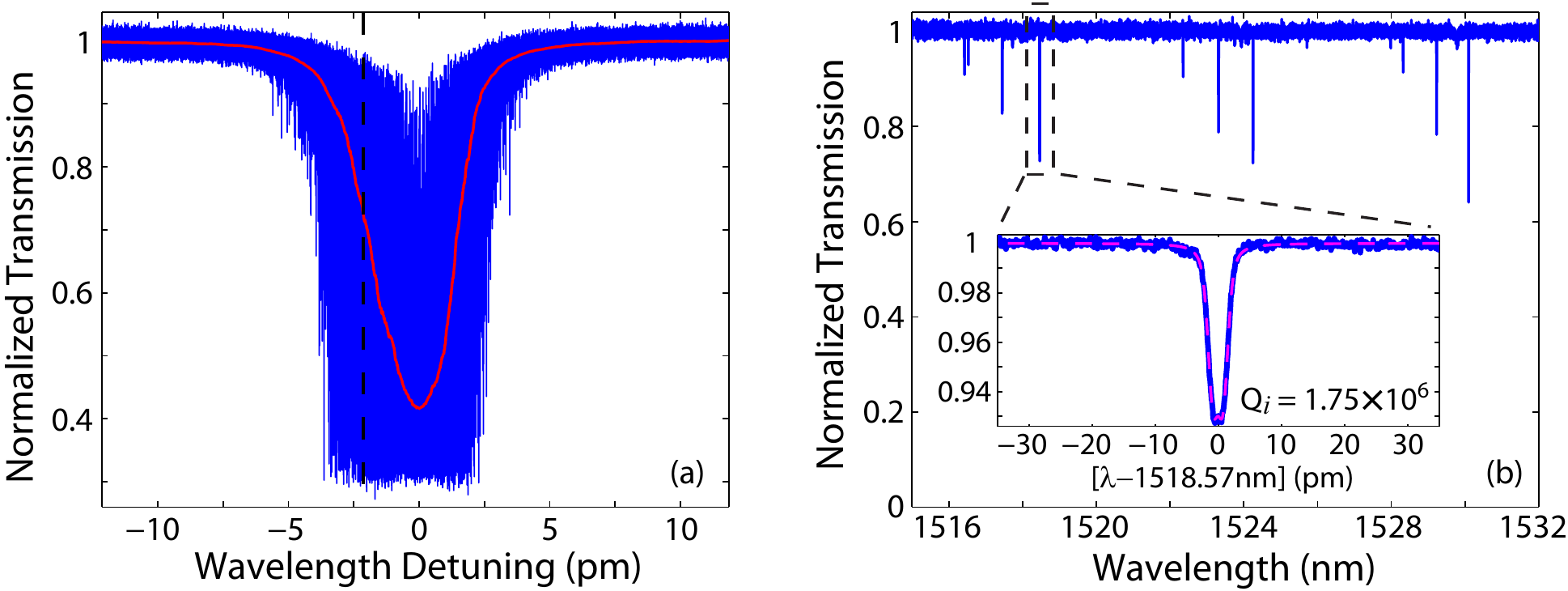}}
\caption{\label{SubFig_CavityTransmission} (a) The cavity transmission of Sample I in a nitrogen environment, when the laser is scanned across the cavity resonance at 1518.57~nm with an input power of 5.8~$\mu$W. The blue curve is the instantaneous signal collected by the high-speed detector and the red curve is the average signal collected by the slow reference detector 2. The slight asymmetry in the transmission spectrum is due to the static component of mechanical actuation when the laser is scanned from blue to red. The dashed line indicates the laser frequency detuning used to record the power spectral density shown in the top panel of Fig.~3(b) in the main text. (b) Linear scan of the averaged cavity transmission of Sample I at an input power of 2.9~${\rm \mu W}$. The inset shows a detailed scan for the bonding mode at 1518.57nm, with the experimental data in blue and the theoretical fitting in red. }
\end{figure}

Unlike other microcavities in which the linear transmission is determined only by the cavity loss and dispersion, for the double-disk NOMS, even the small thermal Brownian motions of the flapping mode introduce significant perturbations to the cavity resonance due to the gigantic opto-mechanical coupling, leading to considerably broadened cavity transmission. Figure \ref{SubFig_CavityTransmission}(a) shows an example of the cavity transmission of Sample I. With a small input power of 5.8 ${\rm \mu}$W well below the oscillation threshold, the cavity transmission exhibits intense fluctuations when the laser frequency is scanned across the cavity resonance. As a result, the averaged spectrum of the cavity transmission (red curve) is significantly broader than the real cavity resonance. A correct description of the cavity transmission requires an appropriate inclusion of the optomechanical effect, which is developed in the following.

When the optical power is well below the oscillation threshold and the flapping mode of the double disk is dominantly driven by thermal fluctuations, the mechanical motion can be described by the following equation:
\begin{equation}
\frac{d^2 x}{dt^2} + \Gamma_{m} \frac{dx}{dt} + \Omega_m^2 x = \frac{F_T(t)}{m_{x}}, \label{FlappingMotion}
\end{equation}
where $\Omega_m$, $\Gamma_{m}$, and $m_{x}$ are the resonance frequency, damping constant, and effective mass of the flapping mode, respectively. $F_T$ is the Langevin force driving the mechanical Brownian motion, a Markovin process with the following correlation function:
\begin{equation}
\ave{F_T(t) F_T(t+\tau)} = 2 m_{x} \Gamma_{m} k_B T \delta (\tau), \label{F_T_Corr}
\end{equation}
where $T$ is the temperature and $k_B$ is Boltzmann's constant. It can be shown easily from Eqs.~(\ref{FlappingMotion}) and (\ref{F_T_Corr}) that the Brownian motion of the flapping mode is also a Markovin process with a spectral correlation given by $\ave{\widetilde{x}(\Omega_1) \widetilde{x}^*(\Omega_2)} = 2 \pi S_x(\Omega_1) \delta(\Omega_1 - \Omega_2)$, where $\widetilde{x}(\Omega)$ is the Fourier transform of the mechanical displacement $x(t)$ defined as $\widetilde{x}(\Omega) = \int_{-\infty}^{+\infty} x(t) e^{i\Omega t} dt$, and $S_x(\Omega)$ is the spectral intensity for the thermal mechanical displacement with the following form:
\begin{equation}
S_x(\Omega) = \frac{2 \Gamma_{m} k_B T/m_{x}}{(\Omega_m^2 - \Omega^2)^2 + (\Omega \Gamma_{m})^2}. \label{S_x}
\end{equation}
The time correlation of the mechanical displacement is thus given by
\begin{equation}
\ave{x(t) x(t+\tau)} = \frac{1}{2 \pi} \int_{-\infty}^{+\infty} S_x(\Omega) e^{-i \Omega \tau} d\tau \equiv \ave{x^2} \rho(\tau) \approx \ave{x^2} e^{-\Gamma_{m} |\tau|/2} \cos \Omega_m \tau , \label{x_Corr}
\end{equation}
where $\ave{x^2}= k_B T/(m_{x} \Omega_m^2)$ is the variance of the thermal mechanical displacement and $\rho(\tau)$ is the normalized autocorrelation function for the mechanical displacement.

To be general, we consider a doublet resonance in which two optical fields, one forward and the other backward propagating, circulate inside the microcavity and couple via Rayleigh scattering from the surface roughness. The optical fields inside the cavity satisfy the following equations:
\begin{eqnarray}
\frac{da_f}{dt} &=& (i \Delta_0 - \kappa/2 - i g_{\text{OM}}x ) a_f + i \eta a_b + i \sqrt{\kappa_{e}} A_{in}, \label{daf_dt}\\
\frac{da_b}{dt} &=& (i \Delta_0 - \kappa/2 - i g_{\text{OM}}x ) a_b + i \eta a_f, \label{dab_dt}
\end{eqnarray}
where $a_f$ and $a_b$ are the forward and backward whispering-gallery modes (WGMs), normalized such that $U_j = |a_j|^2$ ($j=f,b$) represents the mode energy. $A_{in}$ is the input optical wave, normalized such that $P_{in}=|A_{in}|^2$ represents the input power. $\kappa$ is the photon decay rate for the loaded cavity, and $\kappa_{e}$ is the photon escape rate associated with the external coupling. $\Delta_0 = \omega - \omega_0$ is the frequency detuning from the input wave to the cavity resonance and $\eta$ is the mode coupling coefficient. In the case of a continuous-wave input, Eqs.~(\ref{daf_dt}) and (\ref{dab_dt}) provide a formal solution of the forward WGM:
\begin{equation}
a_f(t) = i \sqrt{\kappa_{e}} A_{in} \int_0^{+\infty} {\cos (\eta \tau)  f(\tau) e^{-ig_{\text{OM}} \int_0^\tau {x(t-\tau')d\tau'} }d\tau }, \label{af_t}
\end{equation}
where $f(\tau) \equiv e^{(i\Delta_0-\kappa/2)\tau}$ represents the cavity response. Using Eq.~(\ref{x_Corr}), we find that the statistically averaged intracavity field is given as:
\begin{equation}
\ave{a_f(t)} = i \sqrt{\kappa_{e}} A_{in} \int_0^{+\infty} {\cos(\eta \tau) f(\tau) e^{-\frac{\epsilon }{2} h(\tau)} d\tau }, \label{af_ave}
\end{equation}
where $\epsilon \equiv g_{\text{OM}}^2 \ave{x^2}$ and $h(\tau)$ is defined as
\begin{equation}
h(\tau) \equiv \dint_0^\tau {\rho(\tau_1-\tau_2)d\tau_1 d\tau_2}. \label{h_tau}
\end{equation}
Similarly, we can find the averaged energy for the forward WGM as:
\begin{eqnarray}
\ave{U_f(t)} &=& \kappa_{e} P_{in} \dint_0^{+\infty} {f(\tau_1) f^*(\tau_2) \cos(\eta \tau_1) \cos(\eta \tau_2) e^{ -\frac{\epsilon }{2} h(|\tau_1 - \tau_2|) } d\tau_1 d\tau_2} \nonumber \\
&=& \frac{\kappa_{e} P_{in}}{2 \kappa}  \frac{\kappa - i \eta}{ \kappa - 2 i \eta} \int_0^{+\infty} {e^{-\frac{\epsilon }{2} h(\tau)} \left[ f_c(\tau) + f_s^*(\tau) \right] d\tau} + c.c. , \label{Uf_ave}
\end{eqnarray}
where $f_j(\tau) \equiv e^{(i\Delta_j-\kappa/2)\tau}$ ($j=c,s$), with $\Delta_c = \Delta_0 + \eta$ and $\Delta_s = \Delta_0 - \eta$. $c.c.$ denotes complex conjugate.

As the transmitted power from the double disk is given by
\begin{equation}
P_T(t) = P_{in} + \kappa_{e} U_f(t) + i \sqrt{\kappa_{e}} \left[ A_{in}^* a_f(t) - A_{in} a_f^*(t) \right], \label{P_T}
\end{equation}
the averaged cavity transmission, $\ave{T} \equiv \ave{P_T}/P_{in}$, thus takes the form
\begin{eqnarray}
\ave{T} = 1 - \frac{\kappa_{e} \kappa_{i}}{2 \kappa} \left\{ \left[ 1 - \frac{i \eta \kappa_{e}}{\kappa_{i} (\kappa - 2 i \eta)} \right] \int_0^{+\infty} {e^{ - \frac{\epsilon }{2} h(\tau)} \left[ f_c(\tau) + f_s^*(\tau) \right] d\tau } + c.c. \right\}. \label{T_ave}
\end{eqnarray}
In the case of a singlet resonance, $\eta = 0$ and Eq.~(\ref{T_ave}) reduces to the simple form expression 
\begin{equation}
\ave{T}= 1 - \frac{\kappa_{e} \kappa_{i}}{\kappa} \int_0^{+\infty} {e^{ - \frac{\epsilon }{2} h(\tau)} \left[ f(\tau) + f^*(\tau) \right] d\tau }. \label{T_ave_singlet}
\end{equation}
In the absence of opto-mechanical coupling, $g_{\text{OM}} = 0$ and Eq.~(\ref{T_ave_singlet}) reduces to the conventional form of
\begin{equation}
T = 1- \frac{\kappa_{e} \kappa_{i}}{\Delta_0^2 + (\kappa/2)^2}, \label{T_NoOM}
\end{equation}
as expected.

Using the theory developed above and fitting the experimental averaged cavity transmission spectrum, we obtain the optical Q factor of the resonance, as shown in Fig.~\ref{SubFig_CavityTransmission}(b) for Sample I. The same approach is used to describe the cavity transmission of Sample II, given in Fig.~3(a) of the main text.

\subsection{Power spectral density of the cavity transmission}
\label{appB2}
Here we provide the derivations of the power spectral density of the cavity transmission in the presence of mechanical Brownian motion. We present two theories, one for the linear-perturbation regime when the optomechanical effect is small, the other a non-perturbation theory accurate for arbitrarily strong optomechanical effect.

\subsubsection{The linear-perturbation theory}
\label{appB2i}
If the induced optomechanical perturbations are small, Eq.~(\ref{af_t}) can be approximated as
\begin{equation}
a_f(t) \approx i \sqrt{\kappa_{e}} A_{in} \int_0^{+\infty} {\cos (\eta \tau) f(\tau) \left[ 1 - ig_{\text{OM}} \int_0^\tau {x(t-\tau')d\tau'} \right] }d\tau . \label{af_t_approx}
\end{equation}
In this case, the transmitted optical field can be written as $A_T(t) = A_{in} + i \sqrt{\kappa_{e}} a_f (t) \approx A_0 + \delta A(t) $, where $A_0$ is the transmitted field in the absence of the optomechanical effect and $\delta A$ is the induced perturbation. They take the following forms:
\begin{eqnarray}
A_0 &=& A_{in} \left[ 1 - \kappa_{e} \int_0^{+\infty} {\cos (\eta \tau) f(\tau) d\tau} \right] \equiv A_{in} \hat{A}_0, \label{A_0}\\
\delta A(t) &=& i g_{\text{OM}} \kappa_{e} A_{in} \int_0^{+\infty} {d\tau \cos (\eta \tau) f(\tau) \int_0^\tau {x(t-\tau') d\tau'}}. \label{delta_A}
\end{eqnarray}
The transmitted power then becomes $ P(t) = |A_T(t)|^2 \approx |A_0|^2 + A_0^* \delta A(t) + A_0 \delta A^*(t)$. It is easy to show that $\ave{\delta A (t)} = 0$ and $\ave{P_T(t)} = |A_0|^2$. As a result, the power fluctuations, $\delta P(t) \equiv P_T(t) - \ave{P_T(t)}$, become
\begin{equation}
\delta P(t) \approx g_{\text{OM}} P_{in} \int_0^{+\infty} {d\tau u(\tau) \int_0^\tau {x(t-\tau') d\tau'}}, \label{delta_P}
\end{equation}
where $u(\tau) \equiv i \kappa_{e} \cos (\eta \tau) [\hat{A}_0^* f(\tau) - \hat{A}_0 f^*(\tau)]$. By using Eq.~(\ref{x_Corr}), we find the autocorrelation function for the power fluctuation to be
\begin{equation}
\ave{\delta P(t) \delta P(t+t_0)} \approx \epsilon  P_{in}^2 \dint_0^{+\infty} {d\tau_1 d\tau_2 u(\tau_1) u(\tau_2) \psi (t_0,\tau_1,\tau_2)}, \label{P_Corr}
\end{equation}
where $\psi (t_0,\tau_1,\tau_2)$ is defined as
\begin{equation}
\psi (t_0,\tau_1,\tau_2) \equiv \int_0^{\tau_1} d\tau_1' \int_0^{\tau_2} d\tau_2' \rho(t_0 + \tau_1' - \tau_2'). \label{psi}
\end{equation}
Taking the Fourier transform of Eq.~(\ref{P_Corr}), we obtain the power spectral density $S_P(\Omega)$ of the cavity transmission to be
\begin{equation}
S_P(\Omega) \approx g_{\text{OM}}^2 P_{in}^2 H(\Omega) S_x(\Omega), \label{S_p}
\end{equation}
where $S_x(\Omega)$ is the spectral intensity of the mechanical displacement given in Eq.~(\ref{S_x}) and $H(\Omega)$ is the cavity transfer function given by
\begin{equation}
H(\Omega) = \left| \frac{1}{\Omega} \int_0^{+\infty} {u(\tau) (e^{i\Omega \tau} - 1) d\tau} \right|^2. \label{H}
\end{equation}
In the case of a singlet resonance, the cavity transfer function takes the form:
\begin{equation}
H(\Omega) = \frac{\kappa_{e}^2}{\left[ \Delta_0^2 + (\kappa/2)^2 \right]^2} \frac{4 \Delta_0^2 (\kappa_{i}^2 + \Omega^2)}{\left[ (\Delta_0+\Omega)^2 +(\kappa/2)^2 \right] \left[ (\Delta_0 - \Omega)^2 + (\kappa/2)^2 \right]}. \label{H_Omega_Singlet}
\end{equation}

In most cases, the photon decay rate inside the cavity is much larger than the mechanical damping rate, $\kappa \gg \Gamma_{m}$. For a specific mechanical mode at the frequency $\Omega_m$, the cavity transfer function can be well approximated by $H(\Omega) \approx H(\Omega_m)$. In particular, in the sideband-unresolved regime, the cavity transfer function is given by a simple form of
\begin{equation}
H = \frac{4 \kappa_{e}^2 \kappa_{i}^2 \Delta_0^2}{\left[ \Delta_0^2 + (\kappa/2)^2 \right]^4} . \label{H_Singlet_SidebandUnresolved}
\end{equation}
Therefore, Eq.~(\ref{S_p}) shows clearly that, if the optomechanical effect is small, the power spectral density of the cavity transmission is directly proportional to the spectral intensity of the mechanical displacement.

\subsubsection{The non-perturbation theory}
\label{appB2ii}
The situation becomes quite complicated when the optomechanical effects are large. From Eq.~(\ref{P_T}), the autocorrelation function for the power fluctuation of the cavity transmission, $\delta P(t) \equiv P_T(t) - \ave{P_T}$, is given by
\begin{eqnarray}
\ave{\delta P(t_1) \delta P(t_2)} &=& \kappa_{e}^2 \ave{ U_{f1} U_{f2}} - \kappa_{e} \ave{\left( A_{in}^* a_{f1} - A_{in} a_{f1}^* \right) \left( A_{in}^* a_{f2} - A_{in} a_{f2}^* \right)} \nonumber\\
&+& i \kappa_{e}^{3/2} \left[ \ave{U_{f1}\left( A_{in}^* a_{f2} - A_{in} a_{f2}^* \right) } + \ave{U_{f2} \left( A_{in}^* a_{f1} - A_{in} a_{f1}^* \right)} \right] \nonumber \\
&-& \left[ \kappa_{e} \ave{U_f} + i \sqrt{\kappa_{e}} \left( A_{in}^* \ave{a_f} - A_{in} \ave{a_f^*} \right) \right]^2 , \label{deltaP_Corr}
\end{eqnarray}
where $U_{fj} = U_f(t_j)$ and $a_{fj} = a_f(t_j)$ ($j=1,2$). Equation (\ref{deltaP_Corr}) shows that the autocorrelation function involves various correlations between the intracavity energy and field, all of which can be found using Eqs.~(\ref{x_Corr}) and (\ref{af_t}). For example, we can find the following correlation for the intracavity field:
\begin{eqnarray}
&& \ave{\left( A_{in}^* a_{f1} - A_{in} a_{f1}^* \right) \left( A_{in}^* a_{f2} - A_{in} a_{f2}^* \right)} \nonumber\\
&& = - \kappa_{e} P_{in}^2 \dint_0^{+\infty} {d\tau_1 d\tau_2 C_1 C_2 e^{ - \frac{\epsilon }{2} \left(h_1 + h_2 \right)}
 \left[ f_1 f_2 e^{ - \epsilon  \psi } +  f_1 f_2^* e^{ \epsilon  \psi } + c.c. \right]}, \label{af1af2_ave}
\end{eqnarray}
where, in the integrand, $C_j = \cos (\eta \tau_j)$, $h_j = h(\tau_j)$, $f_j = f(\tau_j)$ (with $j=1,2$), and $\psi = \psi(t_2-t_1, \tau_1, \tau_2)$. $h(\tau)$ and $\psi(t_2-t_1,\tau_1,\tau_2)$ are given by Eqs.~(\ref{h_tau}) and (\ref{psi}), respectively.

Equations (\ref{h_tau}) and (\ref{psi}) show that $h(\tau)$ and $\psi(t_2-t_1,\tau_1,\tau_2)$ vary with time on time scales of $1/\Omega_m$ and $1/\Gamma_{m}$. However, in the sideband-unresolved regime, $\kappa \gg \Gamma_{m}$ and $\kappa \gg \Omega_m$. As the cavity response function $f(\tau)$ decays exponentially with time at a rate of $\kappa/2$, the integrand in Eq.~(\ref{af1af2_ave}) becomes negligible when $\tau_1 \gg 2/\kappa$ or $\tau_2 \gg 2/\kappa$. Therefore, $\psi(t_2-t_1, \tau_1, \tau_2)$ can be well approximated as
\begin{eqnarray}
\psi(t_2-t_1,\tau_1,\tau_2) &=& \frac{1}{2 \pi \ave{x^2}} \int_{-\infty}^{+\infty} {\frac{S_x(\Omega)}{\Omega^2} e^{-i \Omega (t_2 - t_1)} \left(e^{-i\Omega \tau_1} - 1 \right) \left(e^{i\Omega \tau_2} -1 \right) d\Omega } \nonumber \\
& \approx & \frac{\tau_1 \tau_2}{2 \pi \ave{x^2}} \int_{-\infty}^{+\infty} {S_x(\Omega) e^{-i \Omega (t_2 - t_1)} d\Omega } = \tau_1 \tau_2 \rho(t_2-t_1). \label{psi_2}
\end{eqnarray}
Similarly, $h(\tau) \approx \tau^2$, since $h(\tau) = \psi(0,\tau,\tau)$. Therefore, Eq.~(\ref{af1af2_ave}) becomes
\begin{equation}
\ave{\left( A_{in}^* a_{f1} - A_{in} a_{f1}^* \right) \left( A_{in}^* a_{f2} - A_{in} a_{f2}^* \right)} \approx - \kappa_{e} P_{in}^2 \Phi(\Delta t, C_1 C_2), \label{af1af2_ave_approx}
\end{equation}
where $\Delta t = t_2 - t_1$ and $\Phi(\Delta t, C_1 C_2)$ is defined as
\begin{eqnarray}
\Phi(\Delta t, C_1 C_2) \equiv \dint_0^{+\infty} {d\tau_1 d\tau_2 C_1 C_2 e^{ - \frac{\epsilon }{2} \left(\tau_1^2 + \tau_2^2 \right)}
 \left[ f_1 f_2  e^{ - \epsilon  \tau_1 \tau_2 \rho } +  f_1 f_2^* e^{ \epsilon  \tau_1 \tau_2 \rho  } + c.c. \right]}, \quad \label{Phi}
\end{eqnarray}
with $\rho = \rho(\Delta t)$. Following a similar approach, we can find the other correlation terms in Eq.~(\ref{deltaP_Corr}). Using these terms in Eq.~(\ref{deltaP_Corr}), we find that the autocorrelation function of the power fluctuations is given by
\begin{eqnarray}
\ave{\delta P(t_1) \delta P(t_2)} \approx \kappa_{e}^2 P_{in}^2 \Phi(\Delta t, \sigma_1 \sigma_2) - \left[ \kappa_{e} \ave{U_f} + i \sqrt{\kappa_{e}} \left( A_{in}^* \ave{a_f} - A_{in} \ave{a_f^*} \right) \right]^2 , \label{deltaP_Corr_approx}
\end{eqnarray}
where $\sigma_j = \sigma(\tau_j)$ ($j=1,2$) and $\sigma(\tau)$ is defined as
\begin{equation}
\sigma(\tau) \equiv \left[ 1 - \frac{\kappa_{e} (\kappa^2 + 2 \eta^2)}{\kappa (\kappa^2 + 4 \eta^2)} \right] \cos(\eta \tau) + \frac{\eta \kappa_{e}}{\kappa^2 + 4 \eta^2} \sin(\eta \tau). \label{sigma}
\end{equation}

Moreover, Eq.~(\ref{af_ave}) and (\ref{Uf_ave}) show that, in the sideband-unresolved regime, $\ave{a_f}$ and $\ave{U_f}$ are well approximated by
\begin{eqnarray}
\ave{a_f(t)} &\approx& i \sqrt{\kappa_{e}} A_{in} \int_0^{+\infty} {\cos(\eta \tau) f(\tau) e^{-\frac{\epsilon }{2} \tau^2} d\tau }, \label{af_ave_approx}\\
\ave{U_f(t)} &\approx& \kappa_{e} P_{in} \dint_0^{+\infty} {f(\tau_1) f^*(\tau_2) \cos(\eta \tau_1) \cos(\eta \tau_2) e^{ -\frac{\epsilon }{2} (\tau_1 - \tau_2)^2 } d\tau_1 d\tau_2} . \label{Uf_ave_approx}
\end{eqnarray}
Therefore, we obtain the final term in Eq.~(\ref{deltaP_Corr_approx}) as
\begin{equation}
\kappa_{e} \ave{U_f} + i \sqrt{\kappa_{e}} \left( A_{in}^* \ave{a_f} - A_{in} \ave{a_f^*} \right) \approx - \kappa_{e} P_{in} \int_0^{+\infty} {\sigma(\tau) \left[ f(\tau) + f^*(\tau) \right] e^{-\frac{\epsilon}{2} \tau^2} d\tau}. \label{MID1}
\end{equation}
Using this term in Eq.~(\ref{deltaP_Corr_approx}), we obtain the final form for the autocorrelation of the power fluctuations:
\begin{equation}
\ave{\delta P(t_1) \delta P(t_2)} \approx \kappa_{e}^2 P_{in}^2 \left[ \Phi(\Delta t, \sigma_1 \sigma_2) - \Phi(\infty, \sigma_1 \sigma_2) \right] . \label{deltaP_Corr_Final}
\end{equation}
It can be further simplified if we notice that the exponential function $e^{\pm \epsilon \tau_1 \tau_2 \rho(\Delta t)}$ in Eq.~(\ref{Phi}) can be expanded in a Taylor series as
\begin{equation}
e^{\pm \epsilon \tau_1 \tau_2 \rho(\Delta t)} = \sum_{n=0}^{+\infty} {\frac{(\pm \epsilon \tau_1 \tau_2)^n}{n !} \rho^n(\Delta t)}. \label{Taylor}
\end{equation}
Substituting this expression into Eq.~(\ref{Phi}) and using it in Eq.~(\ref{deltaP_Corr_Final}), we obtain the autocorrelation function for the power fluctuation in the following form
\begin{equation}
\ave{\delta P(t) \delta P(t+t_0)} \approx \kappa_{e}^2 P_{in}^2 \sum_{n=1}^{+\infty} {\frac{\epsilon^n \rho^n(t_0)}{n !} \left| G_n^* + (-1)^n G_n \right|^2}, \label{DP_Corr_Sim_Doublet}
\end{equation}
where $G_n$ is defined as
\begin{equation}
G_n \equiv \int_0^{+\infty} {\tau^n \sigma(\tau) f(\tau) e^{-\frac{\epsilon}{2} \tau^2} d\tau}. \label{Gn_Def}
\end{equation}
In the case of a singlet resonance, $\eta=0$ and $\sigma(\tau)$ simplifies considerably to $\sigma = \kappa_{i}/\kappa$. The autocorrelation function for the power fluctuation is still described by Eq.~(\ref{DP_Corr_Sim_Doublet}).

In general, the power spectral density of the cavity transmission is given by the Fourier transform of Eq.~(\ref{DP_Corr_Sim_Doublet}):
\begin{equation}
S_p(\Omega) = \kappa_{e}^2 P_{in}^2 \sum_{n=1}^{+\infty} {\frac{\epsilon^n S_n(\Omega)}{n !} \left| G_n^* + (-1)^n G_n \right|^2}, \label{Sp_nonperturb}
\end{equation}
where $S_n(\Omega)$ is defined as
\begin{equation}
S_n(\Omega) = \int_{-\infty}^{+\infty} {\rho^n (\tau) e^{i\Omega \tau} d\tau}. \label{Sn_nonperturb}
\end{equation}
Eq.~(\ref{S_x}) shows that the spectral intensity of the mechanical displacement can be approximated by a Lorentzian function, resulting in an approximated $\rho(\tau)$ given as $\rho(\tau) \approx e^{-\Gamma_{m} |\tau|/2} \cos \Omega_m \tau$ [see Eq.~(\ref{x_Corr})]. As a result, Eq.~(\ref{Sn_nonperturb}) becomes
\begin{equation}
S_n(\Omega) \approx \frac{1}{2^n} \sum_{k=0}^n {\frac{n!}{k! (n-k)!} \frac{n \Gamma_{m}}{(n\Gamma_{m}/2)^2 + [(2k-n)\Omega_m + \Omega]^2}}. \label{Sn_Sim}
\end{equation}
Combining Eq.~(\ref{Sp_nonperturb}) and (\ref{Sn_Sim}), we can see that, if the optomechanical coupling is significant, the thermal mechanical motion creates spectral components around the harmonics of the mechanical frequency with broader linewidths. As shown clearly in Fig.~3(b) of the main text, the second harmonic is clearly visible. In particular, if the fundamental mechanical linewidth is broad, various frequency components on the power spectrum would smear out, producing a broadband spectral background, as shown in the top panel of Fig.~3(b) in the main text for Sample I. This phenomenon is similar to the random-field-induced spectral broadening in nuclear magnetic resonance \cite{Kubo69} and atomic resonance fluorescence \cite{Kimble77}.

The theory developed in this section can be extended easily for the case with multiple mechanical frequencies. In this case, power spectrum only only exhibits harmonics of each mechanical frequency, but also their frequency sums and differences. As shown in the bottom panel of Fig.~3(b) of the main text, the frequency components near 0 MHz is the differential frequencies and those near 18-20~MHz are the second harmonics and sum frequencies.

\section{Regenerative oscillation and mechanical resonances}
\label{appC}
\subsection{Numerical simulation of optomechanical oscillations}
\label{appC1}
The optomechanical oscillations are simulated through the following coupled equations governing the intracavity optical field and mechanical motions, respectively:
\begin{eqnarray}
&& \frac{da}{dt} = (i \Delta_0 - \frac{\kappa}{2} - i g_{\text{OM}}x ) a + i \sqrt{\kappa_{e}} A_{in}, \label{da_dt}\\
&& \frac{d^2 x}{dt^2} + \Gamma_{m} \frac{dx}{dt} + \Omega_m^2 x = \frac{F_T(t)}{m_{x}} + \frac{F_o(t)}{m_{x}}, \label{dx_dt}
\end{eqnarray}
where we have counted in both the thermal Langevin force $F_T$ and the optical gradient force $F_o = - \frac{g_{\text{OM}} |a|^2}{\omega_0}$ for actuating mechanical motions.

\subsection{Mapping the threshold detuning}
\label{appC2}
\begin{figure}[tbp]
\scalebox{0.440}[0.440]{\includegraphics{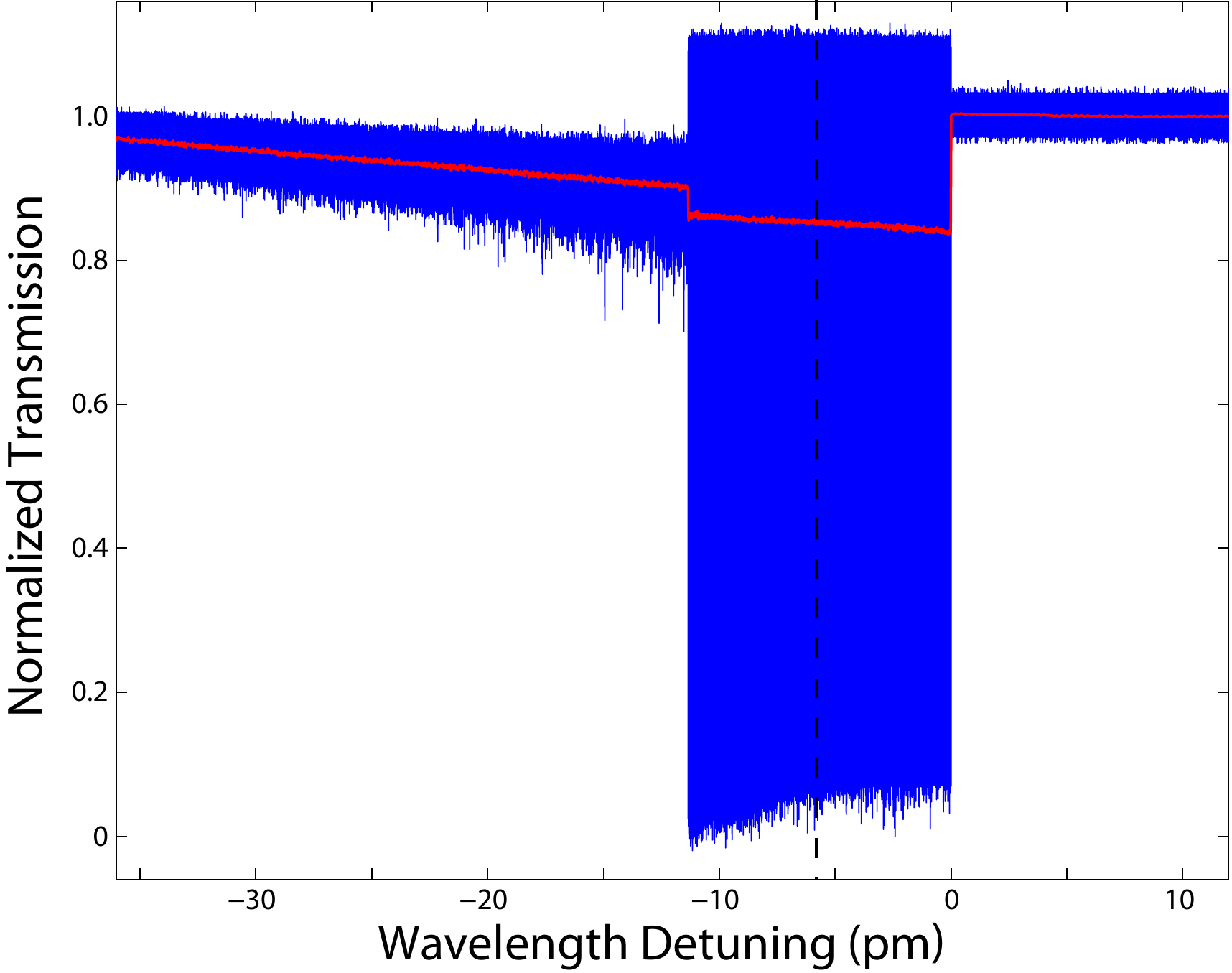}}
\caption{\label{SubFig_DetuningScan1} Scan of the cavity transmission of Sample I at an input power of 0.76 mW, with the instantaneous and averaged signals shown in blue and red, respectively. The dashed line indicated the laser frequency detuning used to record the time-dependent cavity transmission given in Fig.~3(d) in the main text.}
\end{figure}

Figure \ref{SubFig_DetuningScan1} shows an example of the cavity transmission of Sample I. The mechanical flapping mode starts to oscillate when the input laser frequency is scanned across a certain detuning. Within this detuning value, the same magnitude of optomechanical oscillation is excited over a broad range of laser blue detuning. The intense transmission oscillations cover the entire coupling depth, leaving an abrupt kink on the transmission spectrum. The coupling depth at the kink point, $\Delta T_{\rm th}$, corresponds to the threshold coupling at the given power level, from which we can obtain the threshold frequency detuning $\Delta_{\rm th}$.

\subsection{Flapping modes with various azimuthal mode numbers}
\label{appC3}
Because of the extremely short round-trip time of the cavity mode, the optical wave is sensitive only to the variations of averaged disk spacing around the whole disk. As a result, the optomechanical coupling for the fundamental flapping mode, which has flapping amplitude uniformly distributed around the disk perimeter, is maximum, but is nearly zero for flapping mode with high-order azimuthal mode numbers. However, due to the asymmetry in practical devices, the net variations of averaged disk spacing induced by the high-order flapping modes (with azimuthal mode number $\ge$ 1) is not zero, and their thermal motion is visible in the transmission power spectrum. In general, their optomechanical coupling is weak and does not provide efficient dynamic back action.

\section{Cooling of thermal mechanical motion}
\label{appD}
\subsection{Spectral intensity of optically damped thermal mechanical motion}
\label{appD1}
In general, the optomechanical effect is governed by Eqs.~(\ref{da_dt}) and (\ref{dx_dt}). However, the optomechanical effect during mechanical cooling is well described by linear perturbation theory since the thermal mechanical motions are significantly suppressed. The intracavity field can thus be approximated as $a(t) \approx a_0(t) + \delta a(t)$, where $a_0$ is the cavity field in the absence of optomechanical coupling and $\delta a$ is the perturbation induced by the thermal mechanical motions. From Eq.~(\ref{da_dt}), they are found to satisfy the following equations:
\begin{eqnarray}
\frac{da_0}{dt} &=& (i \Delta_0 - \kappa/2) a_0 + i \sqrt{\kappa_{e}} A_{in}, \label{da0_dt}\\
\frac{d \delta a}{dt} &=& (i \Delta_0 - \kappa/2) \delta a - i g_{\text{OM}}x a_0. \label{dDa_dt}
\end{eqnarray}
In the case of a continuous-wave input, Eq.~(\ref{da0_dt}) gives a steady-state value given as:
\begin{equation}
a_0 = \frac{i \sqrt{\kappa_{e}} A_{in}}{\kappa/2 - i \Delta_0},
\end{equation}
and Eq.~(\ref{dDa_dt}) provides the spectral response for the perturbed field amplitude,
\begin{equation}
\delta \widetilde{a}(\Omega) = \frac{i g_{\text{OM}} a_0 \widetilde{x}(\Omega)}{i(\Delta_0 + \Omega) - \kappa/2}, \label{Da_Omega}
\end{equation}
where $\delta \widetilde{a}(\Omega)$ is the Fourier transform of $\delta a(t)$ defined as $\delta \widetilde{a}(\Omega) = \int_{-\infty}^{+\infty} {\delta a(t) e^{i\Omega t} dt}$. Similarly, $\widetilde{x}(\Omega)$ is the Fourier transform of $x(t)$.

The optical gradient force, $F_o = - \frac{g_{\text{OM}} |a|^2}{\omega_0}$, is given by
\begin{equation}
F_o(t) = -\frac{g_{\text{OM}}}{\omega_0} \left[ |a_0|^2 + a_0^* \delta a(t) + a_0 \delta a^*(t)\right]. \label{F_0}
\end{equation}
The first term is a static term which only affects the equilibrium position of the mechanical motion, and can be removed simply by shifting the zero-point of the mechanical displacement to the new equilibrium position. Therefore, we neglect this term in the following discussion. The second and third terms provide the dynamic optomechanical coupling. From Eq.~(\ref{Da_Omega}), the gradient force is given by the following equation in the frequency domain:
\begin{equation}
\widetilde{F}_o(\Omega) = - \frac{2 g_{\text{OM}}^2 |a_0|^2 \Delta_0 \widetilde{x}(\Omega)}{\omega_0} \frac{\Delta_0^2 - \Omega^2 + (\kappa/2)^2 + i \kappa \Omega}{\left[(\Delta_0+\Omega)^2+ (\kappa/2)^2 \right] \left[(\Delta_0 - \Omega)^2 + (\kappa/2)^2 \right]}. \label{F0_Omega}
\end{equation}
As expected, the gradient force is linearly proportional to the thermal mechanical displacement.

Equation (\ref{dx_dt}) can be solved easily in the frequency domain, which becomes
\begin{eqnarray}
(\Omega_{m}^2 - \Omega^2 - i \Gamma_{m} \Omega) \widetilde{x} = \frac{\widetilde{F}_T}{m_{x}} + \frac{\widetilde{F}_o}{m_{x}}. \label{dx_Omega}
\end{eqnarray}
Equation (\ref{dx_Omega}) together with (\ref{F0_Omega}) provides the simple form for the thermal mechanical displacement, 
\begin{equation}
\widetilde{x}(\Omega) = \frac{\widetilde{F}_T}{m_{x}} \frac{1}{(\Omega_{m}')^2 - \Omega^2 - i \Gamma_{m}' \Omega }, \label{dx_Omega2}
\end{equation}
where $\Omega_{m}'$ and $\Gamma_{m}'$ are defined as
\begin{eqnarray}
(\Omega_{m}')^2 &\equiv& \Omega_{m}^2 + \frac{2 g_{\text{OM}}^2 |a_0|^2 \Delta_0}{m_{x} \omega_0} \frac{\Delta_0^2 - \Omega^2 + (\kappa/2)^2}{\left[(\Delta_0+\Omega)^2+ (\kappa/2)^2 \right] \left[(\Delta_0 - \Omega)^2 + (\kappa/2)^2 \right]} \nonumber \\
&\approx& \Omega_{m}^2 + \frac{2 g_{\text{OM}}^2 |a_0|^2 \Delta_0}{m_{x} \omega_0} \frac{\Delta_0^2 - \Omega_{m}^2 + (\kappa/2)^2}{\left[(\Delta_0+\Omega_{m})^2+ (\kappa/2)^2 \right] \left[(\Delta_0 - \Omega_{m})^2 + (\kappa/2)^2 \right]}, \label{Omega_mNew}\\
\Gamma_{m}' &\equiv& \Gamma_{m} - \frac{2 g_{\text{OM}}^2 |a_0|^2 \kappa \Delta_0}{m_{x} \omega_0} \frac{1}{\left[(\Delta_0+\Omega)^2+ (\kappa/2)^2 \right] \left[(\Delta_0 - \Omega)^2 + (\kappa/2)^2 \right]} \nonumber\\
&\approx& \Gamma_{m} - \frac{2 g_{\text{OM}}^2 |a_0|^2 \kappa \Delta_0}{m_{x} \omega_0} \frac{1}{\left[(\Delta_0+\Omega_{m})^2+ (\kappa/2)^2 \right] \left[(\Delta_0 - \Omega_{m})^2 + (\kappa/2)^2 \right]}. \label{Gamma_mNew}
\end{eqnarray}
Equations (\ref{dx_Omega2})-(\ref{Gamma_mNew}) show clearly that the primary effect of the optical gradient force on the mechanical motion is primarily to change its mechanical frequency (the so-called optical spring effect) and energy decay rate to the new values given by Eqs.~(\ref{Omega_mNew}) and (\ref{Gamma_mNew}). The efficiency of optomechanical control is determined by the figure of merit $g_{\text{OM}}^2/m_{x}$. On the red detuned side, the optical wave damps the thermal mechanical motion and thus increases the energy decay rate. At the same time, the mechanical frequency is modified, decreasing with increased cavity energy in the sideband-unresolved regime.

Using Eqs.~(\ref{F_T_Corr}) and (\ref{dx_Omega2}), we find that the spectral intensity of the thermal displacement is given by a form similar to Eq.~(\ref{S_x}):
\begin{equation}
S_x(\Omega) = \frac{2 \Gamma_{m} k_B T/m_{x}}{[(\Omega_m')^2 - \Omega^2]^2 + (\Omega \Gamma_{m}')^2}, \label{S_x2}
\end{equation}
which has a maximum value $S_x(\Omega_m') = \frac{2 \Gamma_{m} k_B T}{m_{x} (\Omega_m' \Gamma_{m}')^2 }$. The variance of the thermal mechanical displacement is equal to the area under the spectrum,
\begin{equation}
\langle (\delta x)^2 \rangle = \frac{1}{2 \pi} \int_{-\infty}^{+\infty}{S_x(\Omega) d\Omega } = \frac{k_B T \Gamma_{m}}{m_{x} (\Omega_m')^2 \Gamma_{m}'}. \label{x_square}
\end{equation}
Cooling the mechanical motion reduces the spectral magnitude and the variance of thermal displacement.

\subsection{Effective temperature of the cooled mechanical mode}
\label{appD2}
For a mechanical mode in thermal equilibrium, the effective temperature can be inferred from the thermal mechanical energy using the equipartition theorem: \begin{equation}
k_B T_{\rm eff} = m_{x} (\Omega_m')^2 \langle (\delta x)^2 \rangle. \label{equipartition}
\end{equation}
The area under the displacement spectrum thus provides an accurate measure of the effective temperature. In practice, fluctuations on the laser frequency detuning may cause the mechanical frequency and damping rate to fluctuate over a certain small range [Eq.~(\ref{Omega_mNew}) and (\ref{Gamma_mNew})], with a probability density function of $p(\Omega_m')$. As a result, the experimentally recorded displacement spectrum is given by the averaged spectrum 
\begin{equation}
\overline{S}_x(\Omega) = \int {S_x(\Omega) p(\Omega_m') d\Omega_m'}, \label{S_x_ave}
\end{equation}
where $S_x(\Omega)$ is given by Eq.~(\ref{S_x2}) and we have assumed $\int {p(\Omega_m') d\Omega_m'} = 1$. The experimentally measured spectral area is thus \begin{equation}
\frac{1}{2\pi} \int_{-\infty}^{+\infty} {\overline{S}_x(\Omega) d\Omega} = \int {\langle (\delta x)^2 \rangle p(\Omega_m') d\Omega_m'} \equiv \overline{\langle (\delta x)^2 \rangle}. \label{x_var_ave}
\end{equation}
Therefore, the integrated spectral area obtained from the experimental spectrum is the averaged variance of thermal mechanical displacement, from which, according to the equipartition theorem, we obtain the effective average temperature
\begin{equation}
k_B \overline{T}_{\rm eff} = m_{x} (\overline{\Omega}_m')^2 \overline{\langle (\delta x)^2 \rangle}, \label{equipartition_ave}
\end{equation}
where $\overline{\Omega}_m' \equiv \int {\Omega_m' p(\Omega_m') d\Omega_m'}$ is the center frequency of the measured displacement spectrum $\overline{S}_x(\Omega)$. Compared with the room temperature, the effective temperature is thus given by
\begin{equation}
\frac{\overline{T}_{\rm eff}}{T_0} = \frac{(\overline{\Omega}_m')^2 \overline{\langle (\delta x)^2 \rangle}}{\overline{\Omega}_m^2 {\langle (\delta x)^2 \rangle}_0}, \label{Tem_ratio}
\end{equation}
where ${\langle (\delta x)^2 \rangle}_0$ is the displacement variance at room temperature, given by the spectral area at $T_0$.

\end{document}